\documentclass[journal,letterpaper]{IEEEtran_nssmic}
\usepackage{cite,epsfig}
\input colordvi.tex

\input babarsym.tex
\def\={\ensuremath{\,}}
\def\cm   {\ensuremath{\rm \,\hbox{cm}}\xspace}
\def\etal{{\em et~al.}}
\def\BaBar{\mbox{\slshape B\kern-0.1em{\smaller A}\kern-0.1em
    B\kern-0.1em{\smaller A\kern-0.1em R}}\xspace}
\def\titlebabar{\mbox{\slshape B\kern-0.1em{\huge A}\kern-0.1em
    B\kern-0.1em{{\huge A}\kern-0.1em {\huge R}}}\xspace}
\def\referencebabar{\mbox{\slshape B\kern-0.1em{\scriptsize A}\kern-0.1em
    B\kern-0.1em{\scriptsize A\kern-0.2em R}}\xspace}

\renewcommand{\gev}{\ensuremath{\mathrm{\,\hbox{Ge}\kern-0.05em\hbox{V}}}\xspace}
\renewcommand{\mev}{\ensuremath{\mathrm{\,\hbox{Me}\kern-0.05em\hbox{V}}}\xspace}

\begin{document}

%%%%%%%%%%%%%%%%%%%%%%%%%%%%%%%%%%%%%%%%%%%%%%%%%%%%%%%%%%%%%%%%%%%%%%%%%%%%%%%%%

{\thispagestyle{empty}

\larger
\onecolumn
\begin{flushright}
physics/0601138 \\
SLAC-PUB-11578 \\
\babar-TALK-05/145 \\
January 2006 \\ 
\end{flushright}

\par\vskip 2.0cm

% Title of the paper
\begin{center}
\Large \bf 
The {\babar} Electromagnetic Calorimeter: \\ Status and
Performance Improvements
\end{center}
\bigskip

\begin{center}
\large Johannes M. Bauer \\ [1mm] for the EMC Group of the \babar\
Collaboration\\ \mbox{ }\\
%\today
\end{center}
\bigskip \bigskip

% Abstract
\begin{center}
\large \bf Abstract
\end{center}\narrower
The electromagnetic calorimeter at the {\babar} detector, part of the
asymmetric B {Factory} at SLAC, measures photons in the energy
range from 20\mev to 8\gev with high resolution.  The~current status of
the calorimeter, now in its seventh year of operation, is being
presented, as well as details on improvements made to the analysis code
during the last years.

\vfill
\begin{center}

Submitted to the Conference Proceedings of the IEEE Nuclear Science
Symposium \\ and Medical Imaging Conference, October 23 -- 29, 2005, Puerto
Rico, U.S.A.

\end{center}

\vspace{1.0cm}
\begin{center}
{\em Stanford Linear Accelerator Center, Stanford University, 
Stanford, CA 94309} \\ \vspace{0.1cm}\hrule\vspace{0.1cm}
Work supported in part by Department of Energy contract DE-AC02-76SF00515 \\
and Department of Energy grant DE-FG05-91ER40622.
\end{center}

}
\newpage
\twocolumn

%%%%%%%%%%%%%%%%%%%%%%%%%%%%%%%%%%%%%%%%%%%%%%%%%%%%%%%%%%%%%%%%%%%%%%%%%%%%%%%%%
%
\title{The \titlebabar Electromagnetic Calorimeter: \\ Status and
Performance Improvements}
\author{Johannes M.~Bauer,~\IEEEmembership{Member,~IEEE,}
        {\em for the EMC Group of the \BaBar Collaboration}% <-this % stops a space
\thanks{Manuscript received November 18, 2005.  This work was supported 
        by U.S. Department of Energy grant DE-FG05-91ER40622.}% <-this % stops a space
\thanks{J.M. Bauer is with the University of Mississippi.}}
% note the % following the last \IEEEmembership and also the first \thanks - 

\maketitle

\begin{abstract}

The electromagnetic calorimeter at the \BaBar detector, part of the
asymmetric {\bf B} {Factory} at SLAC, measures photons in the energy
range from 20\mev to 8\gev with high resolution.  The current status of
the calorimeter, now in its seventh year of operation, is being
presented, as well as details on improvements made to the analysis code
during the last years.

\end{abstract}


\section{The SLAC \B-Factory}

\begin{figure}[bhtp]
\centering
\includegraphics[width=3.5in]{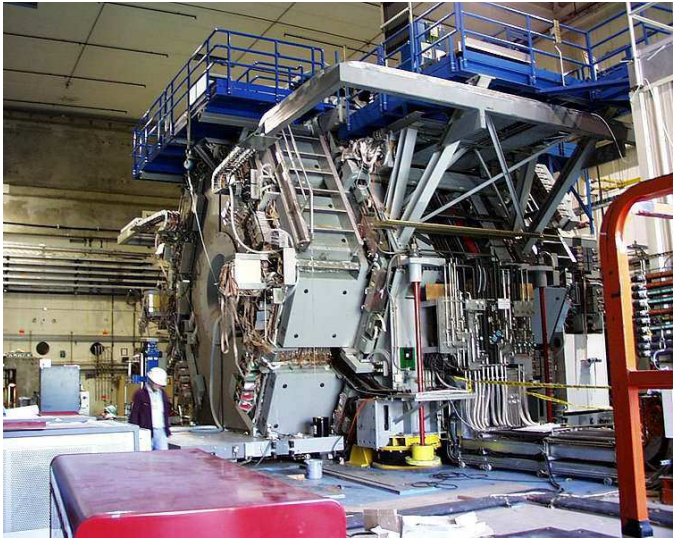}
\caption{The \referencebabar detector during the time of construction.}
\label{BaBarPict}
\end{figure}

\begin{figure}[htbp]
\centering
\includegraphics[width=3.4in]{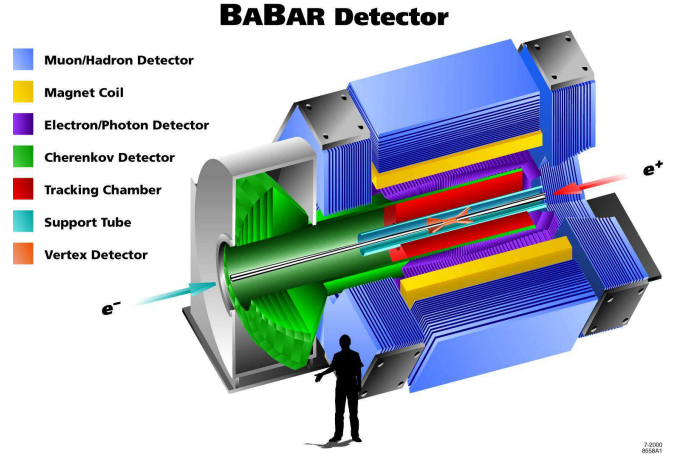}
\caption{Schematic drawing of the \referencebabar detector.  The
         electromagnetic calorimeter is the subsystem shown in purple.}
\label{BaBarDraw}
\end{figure}

\PARstart{S}{ince~1999,} the asymmetric \B factory at the Stanford
Linear Accelerator Center (SLAC) is colliding 9.0\gev electrons with
3.1\gev positrons.  This total energy of 10.58\gev corresponds to the
\FourS resonance.  Its decay particles, \BpBm and \BzBzb pairs, are used
to study \CP violation and many other processes in particle physics.
The~data are collected by the \BaBar Detector (Figs.~\ref{BaBarPict}
and~\ref{BaBarDraw}), which is described in detail in Ref.~\cite{8569}.

\section{Electromagnetic Calorimeter Overview}

\begin{figure}[htbp]
\centering
\includegraphics[width=1.8in]{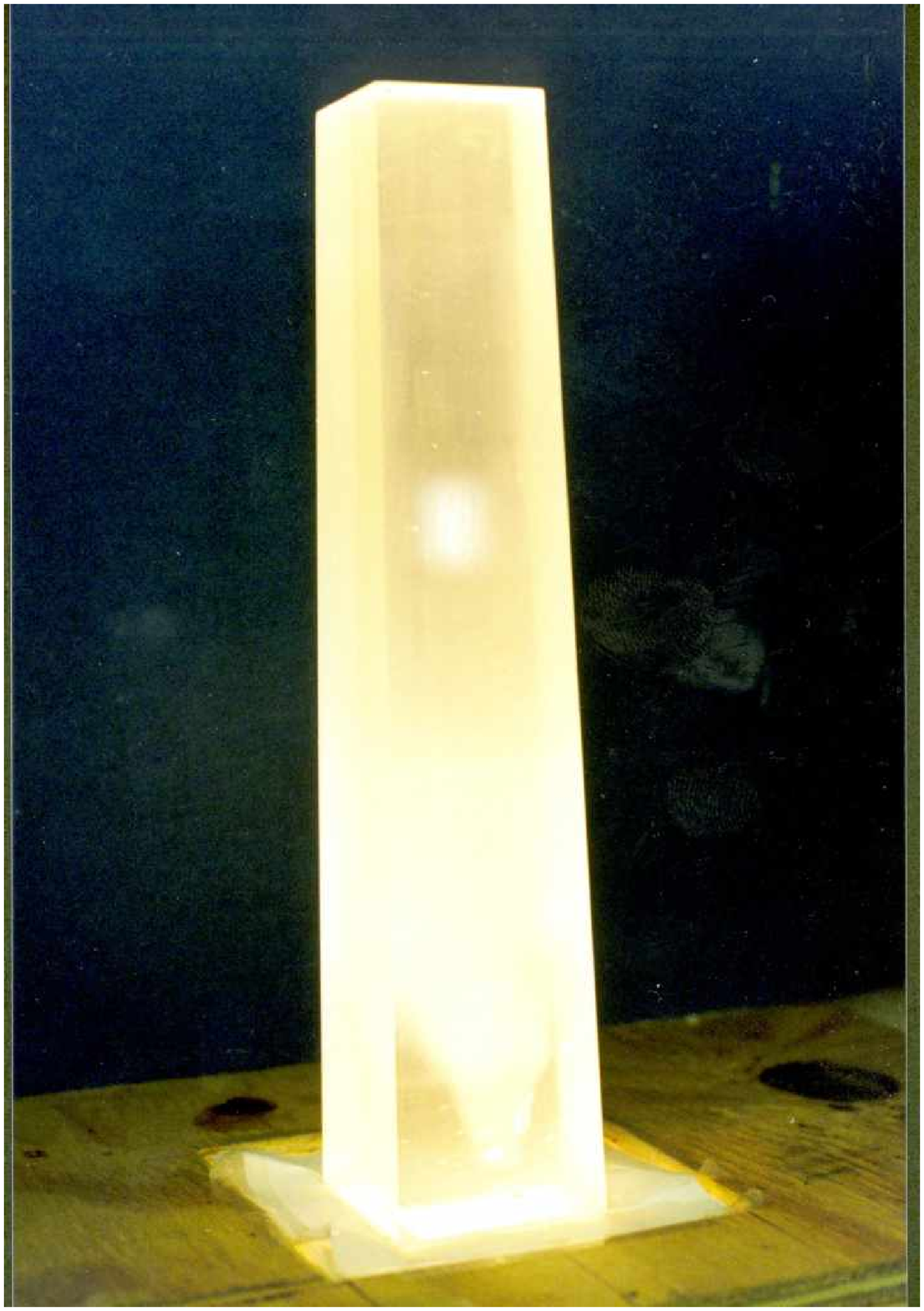} \hskip3mm
\includegraphics[width=1.5in]{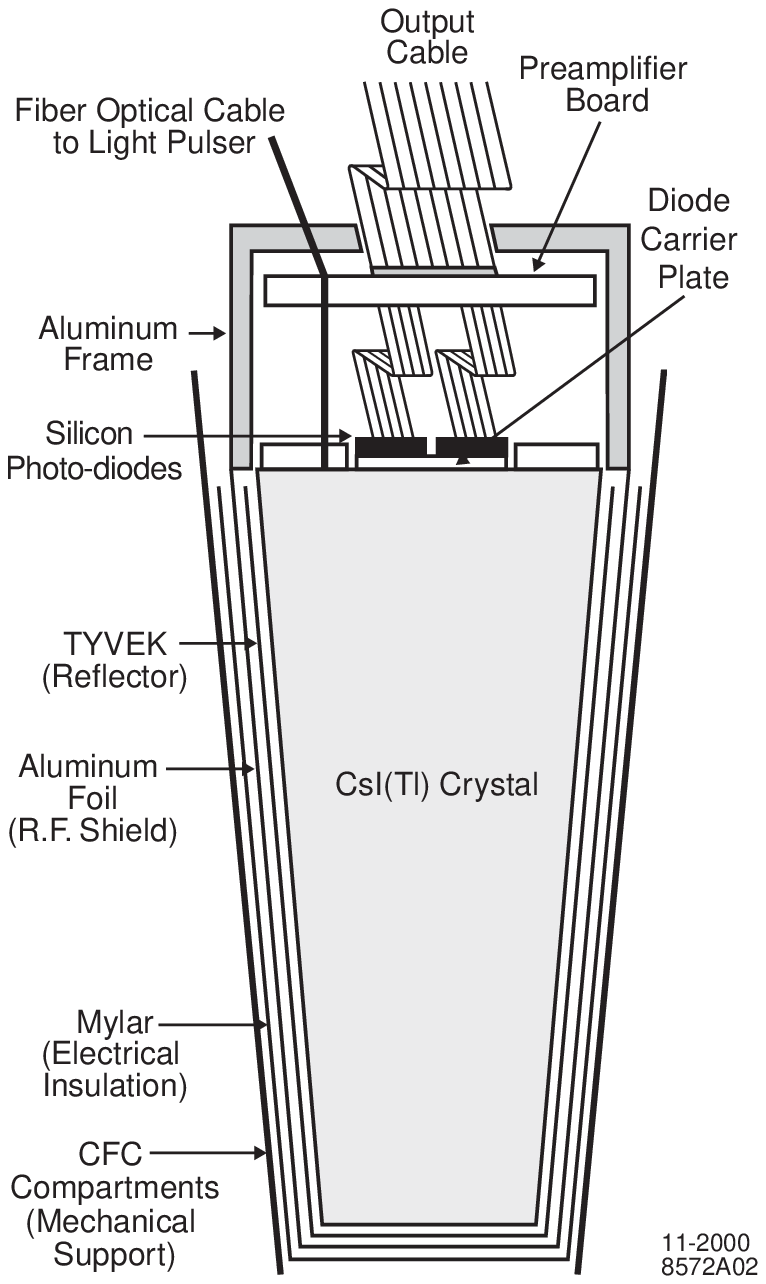}
\caption{Left: Photograph of a CsI(Tl) crystal lit from the bottom by a
         light bulb.  Right: Schematic drawing (not to scale) of a
         crystal with attached electronics.}
\label{Crystal}
\end{figure}

The electromagnetic calorimeter (EMC) consists of 6580 CsI(Tl) crystals
16 to 17.5 radiation lengths deep (Fig.~\ref{Crystal} left), all
pointing close to the interaction point.  At the back of each crystal
two photo diodes and one pre-amplifier card are attached
(Fig.~\ref{Crystal} right).  On~average, the diodes see about 7,300
photo-electrons/\mev.  The~electronics covers the signal with an 18-bit
dynamic range by combining the output of a 10-bit ADC with two range
bits.  This allows the calorimeter to measure photon energies from
20\mev to 8\gev.  The~energy and position resolution was determined to
be the following~\cite{10170}:
\begin{eqnarray}
  \frac{\sigma_E}{E} \hskip-2mm&=&\hskip-2mm \frac{(2.30\pm0.03\pm0.3)\%}{\sqrt[4]{E{\hbox{(GeV)}}}}
    \oplus(1.35\pm0.08\pm0.2)\% \hskip3mm\null\\
  \sigma_\theta \hskip-2mm&=&\hskip-2mm \sigma_\phi = \frac{(4.16\pm0.04)\,\hbox{mrad}}{\sqrt{E{\hbox{(GeV)}}}}
\end{eqnarray}

\begin{figure}[htbp]
\centering
\includegraphics[width=2.3in]{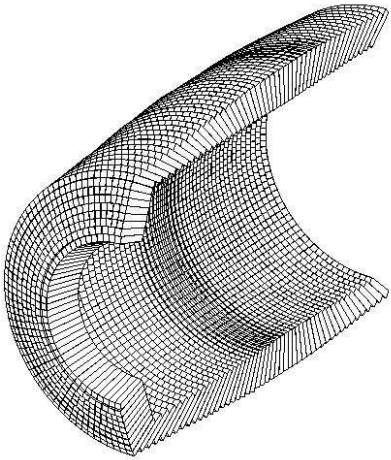}
\vspace{-3mm}
\caption{Cutout drawing of the cylindrical arrangement of the crystals
         into the large barrel and the smaller endcap (left bottom).}
\label{CrystalArrangement}
\end{figure}

\begin{figure}[!t]
\centering
\includegraphics[width=3.5in]{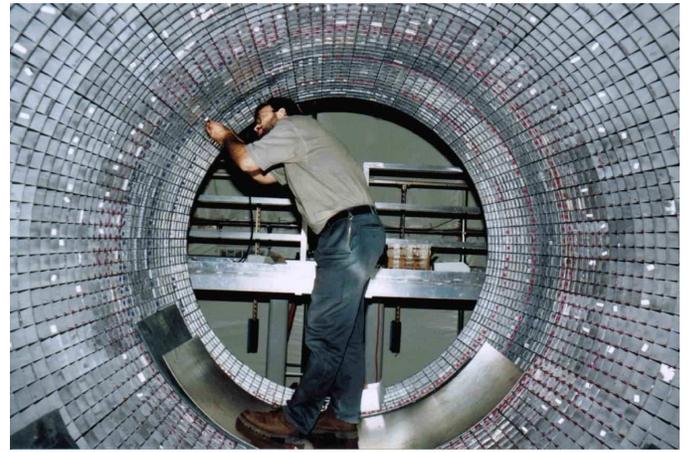}
\caption{Photograph of the inside of the barrel calorimeter during
         construction.}
\label{InsideBarrel}
\end{figure}

\begin{figure}[b!]
\centering
\includegraphics[width=3.5in]{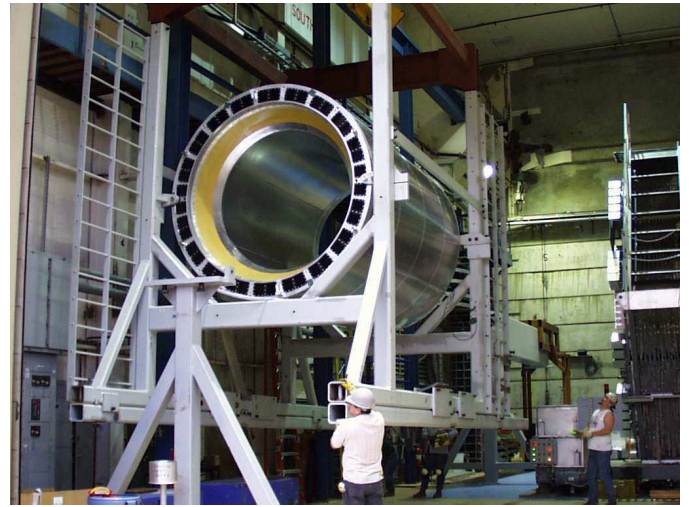}
\caption{Completed barrel calorimeter ready for installation in the
         detector.}
\label{CompletedBarrel}
\end{figure}

The crystals are combined into 7$\times$3 modules (except for 6$\times$3
modules at the backward end and special modules for the forward end),
then assembled like shown in Figure~\ref{CrystalArrangement} into a
barrel and an endcap.  Fig.~\ref{InsideBarrel} gives a view inside the
barrel during assembly, while in Fig.~\ref{CompletedBarrel} the
completed barrel is waiting for its insertion into the detector.

\section{Performance of Hardware}

The operation of the calorimeter is very stable.  Out of 6580 crystals,
only one crystal is dead with no hope for any recovery.  Currently four
more crystals are dead, but they might be recovered at some time.
Fourteen more crystals use only one of the two diodes; several more
crystals are working incorrectly in one energy range, for example at low
energy.  From time to time an ADC board becomes noisy, which, in the
worst case, results in the crystals of this board being masked out until
the board can be replaced during the next access to the detector.  The
electronics is regularly calibrated by determining the pedestals and by
injecting a known charge into the pre-amplifiers to measure the gain and
linearity of the system~(see also Ref.~\cite{ivo}).

\section{Calibration of Individual Crystals}

The individual response of a crystal to deposited energy, namely its
overall light yield and its non-uniformity in the light output (the
dependence on the location of the shower inside the crystal), depends on
the details at time of manufacture and the shape of the crystal.  The
light output also decreases over time due to radiation damage.  Each 
crystal of the \BaBar calorimeter is therefore regularly
calibrated.  Two absolute energy calibrations are employed for this: The
liquid source calibration at low energy and the Bhabha calibration at
high energy.  For intermediate energies the calibration constants are
interpolations between these two extremes following a function linear in
the logarithm of the energy.

\subsection{Liquid Source System}

\begin{figure}[htbp]
\centering
\includegraphics[width=2.2in]{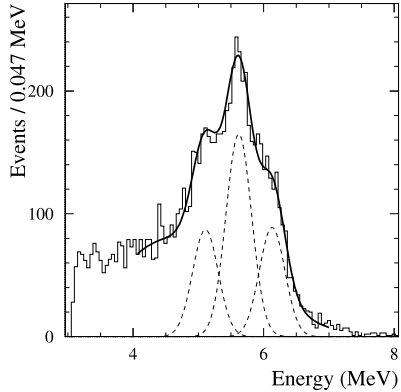}
\vspace{-6mm}
\caption{Spectrum of 6.13\mev photons as detected by a crystal of the
         calorimeter.  The~Gaussian functions indicate the contributions
         from the 6.13\mev peak (far right) and the two escape peaks
         (middle and left).}
\label{LSspectrum}
\end{figure}

\begin{figure}[htbp]
\centering
\vspace{-5mm}
\includegraphics[width=3.7in]{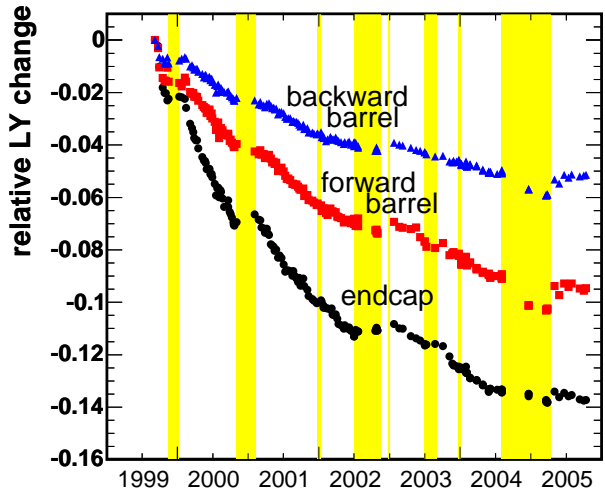}
\vspace{-18mm}
\caption{Relative loss in light yield over time for the backward barrel
         (blue), forward barrel (red) and endcap (black) as measured by
         the liquid source system.  The yellow areas indicate major
         times without beam.}
\label{LSloss}
\end{figure}

Whenever a liquid source calibration is performed, a neutron generator
is switched on to emit 14\mev neutrons.  The generator is surrounded by
Fluorinert (FC77), a liquid rich in fluorine, and the~following chain
results in the emission of 6.13\mev photons through the decay of
$^{16}$N with a half-life time of 7\=seconds:
\begin{equation}
  ^{19}\hbox{F} + \hbox{n} \to ^{16}\hbox{N} + \alpha \end{equation}
\begin{equation}
  ^{16}\hbox{N} \to  \hbox{$^{16}$O}^* \to \hbox{$^{16}$O} + \g (6.13\mev) 
\end{equation}
A~system of pipes transports the radioactive liquid past the front of
the crystals.  There the photons enter the crystals and are detected
with the regular data acquisition system.  Figure~\ref{LSspectrum} shows
the spectrum of these photons as seen by a crystal in the calorimeter.
Calibrations are performed about once a month to a statistical
uncertainty of $\le$0.5\% and a systematic uncertainty of about 0.1\%.
The~average loss in light yield over time due to radiation damage as
measured by the liquid source system is documented in Fig.~\ref{LSloss}.
Radiation measurements by RadFETs located at the calorimeter indicate
that the average radiation dose so far is $\sim$0.7\=kRad for the barrel
and $\sim$1.1\=kRad for the endcap\cite{RadFET}\cite{Jong}.  For more
details on the liquid source calibration system see Ref.~\cite{10289}.

\subsection{Bhabha Calibration}

\begin{figure}[htbp]
\centering
\includegraphics[width=3.5in]{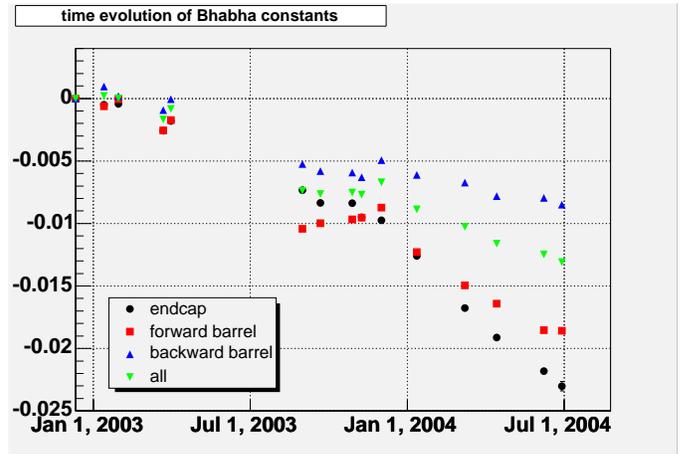}
\vspace{-4mm}
\caption{Evolution of average Bhabha constants over time.}
\label{bhabha}
\end{figure}

The second absolute energy calibration of individual crystals is
performed with $\ep\en\to\ep\en$ events from regular recorded data.
These calibrations involve crystal energies of 2.5\gev to 8\gev,
depending on the polar angle due to the asymmetry in the beam energies.
The~calibration requires most crystals to have at least 200 direct hits
in order to reach a statistical error of 0.35\% for individual crystals.
The systematic error is estimated to be less than 1\%~\cite{ralph}.
Calibration constants are currently calculated up to once a month, but
will soon be obtained more frequently once the code is running
automatically as part of the regular reconstruction system.
The~constants change over time in a way similar to the source
calibration constants (Fig.~\ref{bhabha}).

\section{Cluster Calibrations}

Not only calibrations of individual crystals are needed, but also
calibrations of the clusters, which are groups of adjacent crystals in
which the full shower energy of a particle is deposited.  These
corrections adjust for shower energy lost at the rear of the crystals,
gaps between the crystals, and the sides of the calorimeter.

\begin{figure}[htbp]
\centering
\includegraphics[width=3.5in]{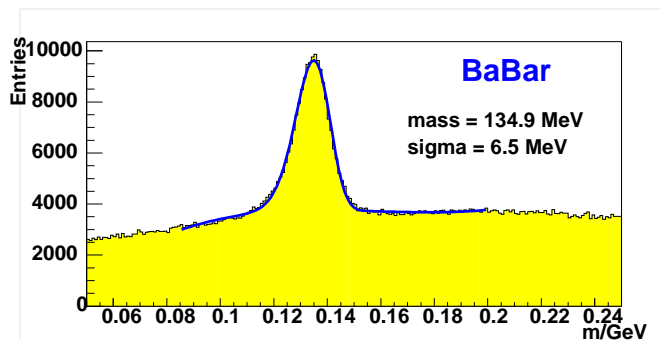}
\vspace{-4mm}
\caption{Distribution of the invariant mass of photon pairs after
         applying the cluster calibration.}
\label{pi0peak}
\end{figure}

For~clusters with energies up to 2\gev, the calibration is obtained from
\piz mesons by correcting the photon energies so that the distribution
of the invariant mass of two photons agrees with the expected $\piz$
mass distribution.  Corrections are mostly in the 6\% to 8\% range.
Figure~\ref{pi0peak} shows a typical $m(\g\g)$ distribution with a clear
\piz peak.  An improved version of the calibration is currently being
tested.

For clusters with energies above 2\gev, the correction factors are
obtained from single-photon Monte Carlo simulation.  Soon new
calibration constants based on $\ep\en\to\mu\mu\gamma$ events will be
introduced.

\section{Improvements in Reconstruction Software}

\subsection{Position of Cluster Inside Crystals}

\begin{figure}[htbp]
\centering
\includegraphics[width=2.8in]{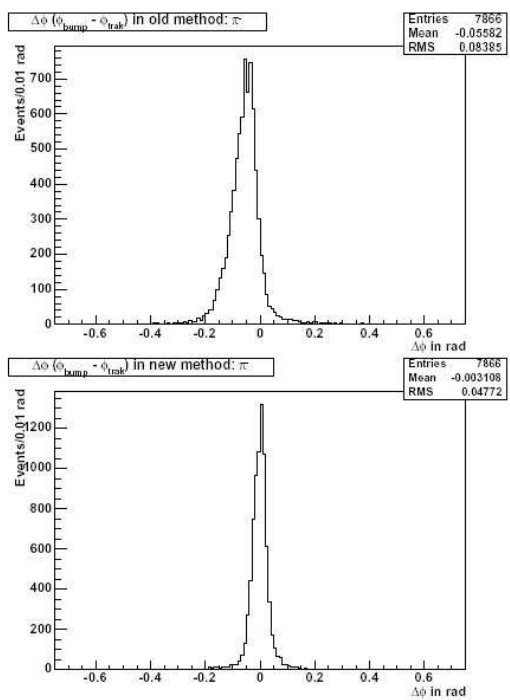}
\vspace{-3mm}
\caption{Distribution of the azimuthal angle difference (in rad) of the
         cluster position and the point where the track intersects with
         the calorimeter.  The~plots are based on actual data with
         requirements applied that select a clean set of $\pi$ mesons.
         For the top plot the old, for the bottom plot the new cluster
         position algorithm is used.}
\label{matchPhi}
\end{figure}

\begin{figure}[htbp]
\centering
\includegraphics[width=3.4in]{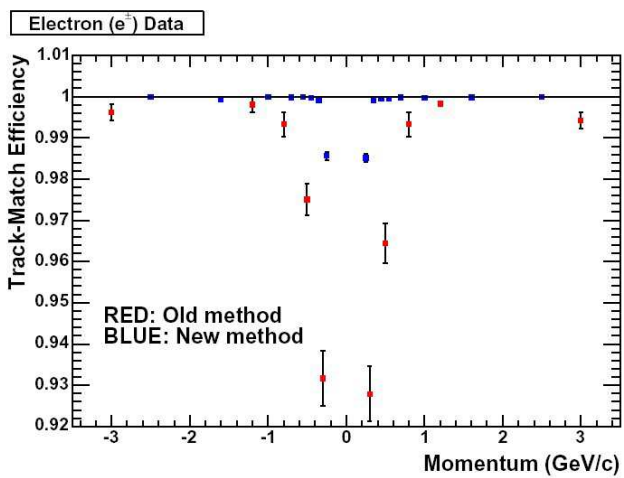}
\vspace{-4mm}
\caption{Track matching efficiency versus momentum for electrons and
         positrons.  The plot is based on actual data with requirements
         that select a clean set of electrons and positrons.  For the
         points in red the old, for the points in blue the new cluster
         position algorithm is used.}
\label{matchEff}
\end{figure}

When matching a track to a cluster of the calorimeter, the position of
the cluster in three dimensions has to be known.  Until recently, this
position of the cluster center was always located at the front of the
crystals.  This caused less than optimal performance of the matching
algorithm, such as when a low-energy track, spiraling in the magnetic
field of the detector, enters the calorimeter at an~angle.  Now all
clusters are assumed to be located at a depth of 12.5\cm inside the
crystals.  Figure~\ref{matchPhi} displays the improvement in the
matching by showing the azimuthal angle difference of the matched tracks
and clusters before and after this change.  Similarly, the
track-matching efficiency improved, especially at very low momentum
(Fig.~\ref{matchEff}).

\subsection{Edge Correction}

\begin{figure}[htbp]
\centering
\includegraphics[width=3.5in]{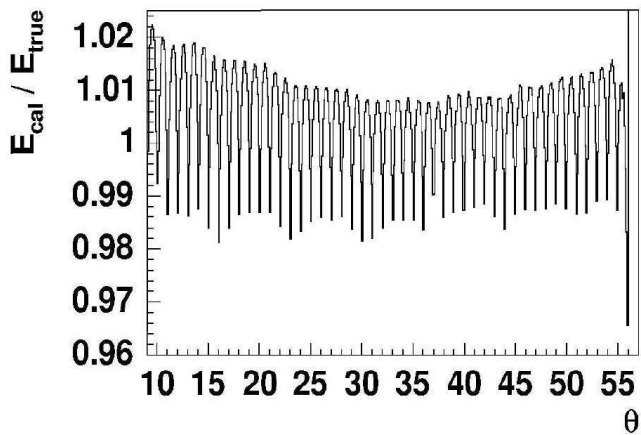}
\vspace{-9mm}
\caption{Ratio of reconstructed over true energy of photons versus
         crystal ring of the barrel.  The forward end (where the endcap
         is attached) is on the left side, the backward end on the right
         side.  The plot is based on \BB Monte Carlo simulation with
         measured photon energy $E_{\rm cal}$ between 0.5\gev
         and~0.8\gev.}
\label{edgeTheta}
\end{figure}

\begin{figure}[htbp]
\centering
\includegraphics[width=3.5in]{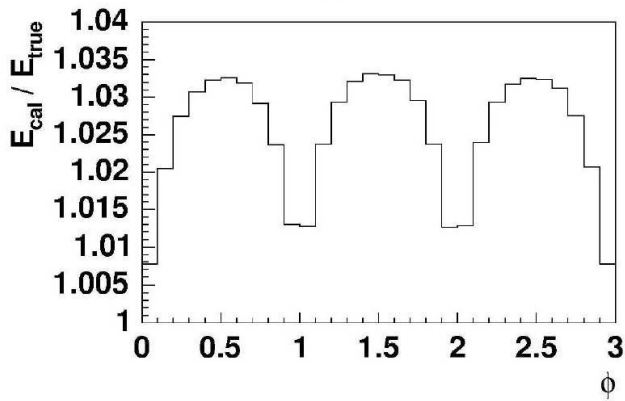}
\vspace{-7mm}
\caption{Ratio of reconstructed over true energy of photons versus
         azimuthal angle in units of crystals.  Due to the modular
         structure in the calorimeter, the crystals are folded over into
         one single group of three crystals.  The dips at 0 and 3 are
         deeper because of the larger gap between the crystal modules.
         The plot is based on \BB Monte Carlo simulation with measured
         photon energy $E_{\rm cal}$ between 0.5\gev and~0.8\gev.}
\label{edgePhi}
\end{figure}

\begin{figure}[htbp]
\centering
\includegraphics[width=3.5in]{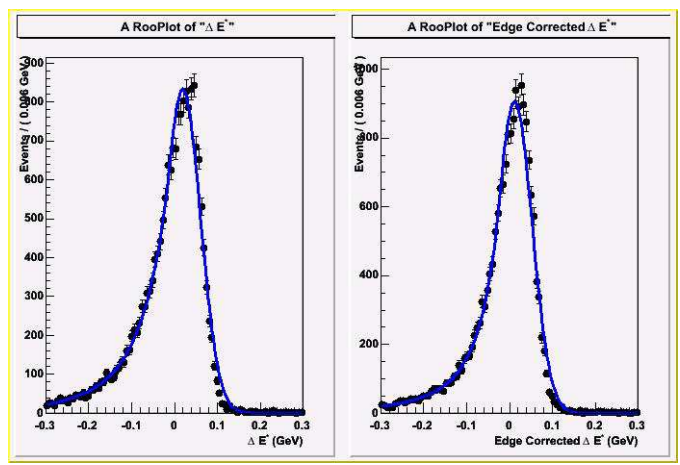}
\caption{Distribution of the quantity \DeltaE for signal Monte Carlo
         events $\Bp\to\Kstarp\g$.  The~left plot was obtained without,
         the right plot with the edge correction.}
\label{Kstarg}
\end{figure}

If~a photon hits the calorimeter at a position close to the edge between
two crystals, up to 3\% of its energy is lost in gaps, as can be seen
from Figs.~\ref{edgeTheta} and~\ref{edgePhi}.  A~correction is now
applied to the energy of each cluster.  The effect of this so-called
``edge correction'' on physics analyses can be seen from the
distributions in Fig.~\ref{Kstarg}.  The~quantity plotted is \DeltaE,
the difference between the measured energy of a \B meson candidate minus
the known beam energy.  Without measurement uncertainties, the peak
would be exactly at zero.  The~underlying data set of the shown
distributions is Monte Carlo simulation of the decay $\Bp\to\Kstarp\g$.
The~left plot, obtained without edge correction, has a FWHM/2.36 of
($45.1\pm0.7$)\mev, while the right plot, obtained with edge correction,
has a FWHM/2.36 of ($42.0\pm0.6$)\mev, which means that in this case the
edge correction improved the \DeltaE resolution by 7\%.

\section{Additional Studies and Future Goals}

Many decay modes are being used to study the performance of the
calorimeter, like $\ep\en\to\mu\mu\g$ events, radiative Bhabha events
($\ep\en\to\ep\en\g$), $\ep\en\to\g\g$, $\Dstarz\to\Dz\g$ (photon
energies from 100 to 400\mev), $\Sigma_0\to\Lambda\g$ (photon energies
from 50 to 250\mev).  A~new cluster calibration is about to be
implemented, and the Bhabha calibration will soon be automated to
provide more frequent monitoring and correction of the radiation damage
at high energies.

\section{Conclusion}

The \BaBar electromagnetic calorimeter operates very reliably and
delivers very good performance for the experiment.  The~damage to the
crystals due to radiation is regularly measured and calibrated out.
Over time, enhancements were added to the reconstruction code, and the
tweaking of the calibrations continues in order to improve the
reconstruction of the detected particles and ultimately improve the
physics analyses of \BaBar.

\section*{Acknowledgment}
The author thanks everyone in the \BaBar EMC group for all their
contributions to the calorimeter.  He~congratulates and extends his
gratitude to the whole \BaBar Collaboration and the PEP-II accelerator
group for their tremendous accomplishments.

\newpage

\end{document}